\documentclass{jnmp}

\usepackage{amsmath}

\def\tr{\mbox{tr}\,}

\def\Ad{\mbox{Ad}\,}
\def\Aut{\mbox{Aut}\,}
\def\fr#1{{\mathfrak{#1}}}
\def\openone{\leavevmode\hbox{\small1\kern-3.3pt\normalsize1}}

\JNMPnumberwithin{equation}{section}

\theoremstyle{definition}

\begin{document}

%
\renewcommand{\evenhead}{G G Grahovski and M Condon}
\renewcommand{\oddhead}{On the Caudrey-Beals-Coifman System and the Gauge Group Action}

%
\thispagestyle{empty}

\FirstPageHead{*}{*}{2008}{\pageref{firstpage}--\pageref{lastpage}}{Article}

\copyrightnote{2007}{G G Grahovski and M Condon}

\Name{On the Caudrey-Beals-Coifman System and the Gauge Group Action}

\label{firstpage}

\Author{Georgi G. GRAHOVSKI~$^{\dag,\ddag}$ and Marissa CONDON~$^\ddag$}

\Address{$^\dag$ Institute for Nuclear
Research and Nuclear Energy, Bulgarian Academy of Sciences, 72 Tsarigradsko chauss\'ee, 1784 Sofia,
Bulgaria \\
~~E-mail: grah@inrne.bas.bg\\[10pt]
$^\ddag$ School of
Electronic Engineering, Dublin City University, Glasnevin, Dublin
9, Ireland \\
~~E-mail: grah@eeng.dcu.ie, \qquad condonm@eeng.dcu.ie}

\Date{Received Month *, 200*; Revised Month *, 200*; 
Accepted Month *, 200*}

\begin{abstract}
\noindent
The generalized Zakharov--Shabat
systems with complex-valued Cartan elements and the systems
studied by Caudrey, Beals and Coifman (CBC systems) and
their gauge equivalent are studies. This includes:
the
properties of fundamental analytical solutions (FAS)
for the gauge-equivalent to CBC systems and the minimal set of
scattering data; the description of the class of
nonlinear evolutionary equations solvable by the inverse
scattering method and the recursion operator, related to such
systems; the hierarchies of Hamiltonian structures.
\end{abstract}

%
\section{Introduction}\label{sec:1}

The idea that the inverse scattering method (ISM) is a generalized Fourier
transform has appeared as early as 1974 in \cite{AKNS}. In the class of
nonlinear evolution equations (NLEE) related to the Zakharov--Shabat (ZS)
system \cite{Za*Sh,1} Lax operator belonging to ${\it sl(2)} $ algebra was studied.
This class of NLEE
contains such physically important equations as
the nonlinear Schr\"odinger equation (NLS), the sin-Gordon and modified
Korteveg--de-Vriez (mKdV) equations.

The multi-component ZS system leads to such important systems
like the multi-com\-po\-nent NLS, the $N$-wave type equations, etc.

Here we consider the $n \times n $ system
\cite{BS,Caud*80,ForGib*80b}:
\begin{eqnarray}\label{eq:1.5} L\Psi
(x,t,\lambda )= \left({\rm i}{{\rm d} \over {\rm d} x}+q(x,t)-\lambda J\right) \Psi
(x,t,\lambda ),
\end{eqnarray} where $q(x,t) $ and $J $ take values in the
semi-simple Lie algebra ${\frak g} $
\cite{MiOlPer*81,G*86,Za*Mi,ForKu*83}:
\begin{eqnarray}\label{eq:1.6}
q(x,t) = \sum_{\alpha \in \Delta _+} \left(q_{\alpha }(x,t)E_{\alpha }+
q_{-\alpha }(x,t)E_{-\alpha } \right)\in {\frak g}_J \qquad
J = \sum_{j=1}^{r}a_jH_j \in {\frak h}. \nonumber
\end{eqnarray}
For the case of complex $J$ we will refer this system as Caudrey-Beals-Coifman (CBC) system. 
Here $J $ is a regular element in the Cartan subalgebra ${\frak
h} $ of ${\frak g} $, ${\frak g}_J $ is the image of $\mbox{ad}_J
$, $\{E_{\alpha }, H_i\} $ form the Cartan--Weyl Basis in ${\frak
g} $, $\Delta _+ $ is the set of positive roots of the algerbra,
$r= \mbox{rank}\, {\frak g}= \mbox{dim}\, {\frak h}$. For more
details see section 2 below. The regularity of the Cartan
elements means that ${\frak g}_J $ is spanned by all root vectors
$E_{\alpha } $ of ${\frak g} $, i.e. $\alpha (J) \neq 0 $ for any
root $\alpha  $ of ${\frak g} $.

The given NLEE as well as the other members of its hierarchy posses
Lax representation of the form (according to (\ref{eq:1.5})):
$
[L(\lambda ), M_P(\lambda )]=0,
$
where
\begin{eqnarray}\label{eq:1.7}
M_P\Psi (x,t,\lambda ) = \left({\rm i}{{\rm d} \over {\rm d} t}+\sum_{k=-S}^{P-1}
V_k(x,t)-\lambda ^Pf_PI \right) \Psi (x,t,\lambda )=0, \qquad
I\in {\frak h},
\end{eqnarray}
which must hold identically with respect to $\lambda  $. A
standard procedure generalizing the AKNS one \cite{AKNS} allows
us to evaluate $V_k(x,t) $ in terms of $q(x,t) $ and its
$x$-derivatives. Here and below, we consider only the class of
potentials $q(x,t) $ vanishing fast enough for $|x|\rightarrow
\infty  $. Then one may also check that the asymptotic value of
the potential in $M_P(\lambda ) $ namely $f^{(P)}(\lambda
)=f_P\lambda ^PI $ may be understood as the dispersion law of the
corresponding NLEE. 

Another important trend in the development of IST was the introduction of the 
reduction group by A. V. Mikhailov \cite{2}, and
further developed in \cite{ForGib*80b,ForKu*83,Za*Mi,MiOlPer*81,Ivanov}.
This allows one to prove that some of the well
known models in the field theory \cite{2} and also a number of new
interesting NLEE \cite{2,ForGib*80b,MiOlPer*81} are integrable by
the ISM and posses special symmetry properties. As a result its
potential $q(x,t) $ has a very special form and J can no-longer
be chosen real.

This problem of constructing the spectral theory for (\ref{eq:1.5}) in the
most general case when $J $ has an arbitrary complex eigenvalues was
initialized by Beals, Coifman and Caudrey \cite{BC,BC2,BC3,Caud*80} and
continued by Zhou \cite{Zhou} in the case when the algebra ${\frak g} $ is $
sl(n) $, $q(x,t) $ vanishing fast enough for $|x| \rightarrow \infty  $
and no a priori symmetry conditions are imposed on $q(x,t) $. This has
been done later for any semi-simple Lie algebras by
Gerdjikov and Yanovski \cite{VYa}. 

The zero-curvature condition $
[L(\lambda ), M_P(\lambda )]=0,
$ is invariant under
the action of the group of gauge transformations \cite{ZaTa}.
Therefore the gauge equivalent systems are again completely
integrable, posses hierarchy of Hamiltonian structures, etc,
\cite{FaTa,1,VYa,ZaTa}.

The structure of this paper is as follows: In section 2 
we summarize some basic facts about the reduction
group and Lie algebraic details. The construction of the fundamental analytic solutions (FAS) is
sketched in section 3 which is done separately for the case of
real Cartan elements (section 3.1) and for complex ones (section
3.2). The gauge equivalent NLEE's to the CBC systems are described in section 4.

\section{Preliminaries}\label{sec:2}

\subsection{Simple Lie Algebras}\label{ssec:2.1}

Here we fix up the notations and the normalization conditions for the
Cartan-Weyl generators of ${\frak g} $ \cite{Helg}. We introduce $h_k\in
{\frak h} $, $k=1,\dots,r $ and $E_\alpha  $, $\alpha \in \Delta  $ where
$\{h_k\} $ are the Cartan elements dual to the orthonormal basis $\{e_k\}$
in the root space ${\mathbb E}^r $. Along with $h_k $, we introduce also
\begin{equation}\label{eq:31.1}
H_\alpha = {2  \over (\alpha ,\alpha ) } \sum_{k=1}^{r} (\alpha ,e_k) h_k,
\quad \alpha \in \Delta ,
\end{equation}
where $(\alpha ,e_k) $ is the scalar product in the root space
${\mathbb E}^r $ between the root $\alpha  $ and $e_k $. The commutation
relations are given by:
\begin{eqnarray*}\label{eq:31.2} [h_k,E_\alpha
] = (\alpha ,e_k) E_\alpha , \quad [E_\alpha ,E_{-\alpha }]=H_\alpha ,
\quad [E_\alpha ,E_\beta ] = \left\{ \begin{array}{ll} N_{\alpha
,\beta } E_{\alpha +\beta } \quad & \mbox{for}\; \alpha +\beta \in \Delta
\\ 0 & \mbox{for}\; \alpha +\beta \not\in \Delta \cup\{0\}. \end{array}
\right.  \end{eqnarray*}

We will denote by $\vec{a}=\sum_{k=1}^{r}a_k e_k $ the $r $-dimensional
vector dual to $J\in {\frak h} $; obviously $J=\sum_{k=1}^{r}a_k h_k $. If $
J  $ is a regular real element in ${\frak h} $ then without restrictions we
may  use it to introduce an ordering in $\Delta  $. Namely we will
say that the root $\alpha \in\Delta _+ $ is positive (negative) if
$(\alpha ,\vec{a})>0 $ ($(\alpha ,\vec{a})<0 $ respectively).
The normalization of the basis is determined by:
\begin{eqnarray}\label{eq:32.1}
E_{-\alpha } =E_\alpha ^T, \quad \langle E_{-\alpha },E_\alpha \rangle
={2  \over (\alpha ,\alpha ) }, \quad
N_{-\alpha ,-\beta } = -N_{\alpha ,\beta }, \quad N_{\alpha ,\beta } =
\pm (p+1),
\end{eqnarray}
where the integer $p\geq 0 $ is such that $\alpha +s\beta \in\Delta  $ for
all $s=1,\dots,p $ $ \alpha +(p+1)\beta \not\in\Delta  $ and $\langle
\cdot,\cdot \rangle  $ is the Killing form of ${\frak  g} $.  The root
system $\Delta $ of ${\frak g} $ is invariant with respect to the Weyl
reflections $A^*_\alpha $; on the vectors $\vec{y}\in {\mathbb E}^r $ they
act
as
$
A^*_\alpha \vec{y} = \vec{y} - {2(\alpha
,\vec{y})  \over (\alpha ,\alpha )} \alpha , \quad \alpha \in \Delta$.
All Weyl reflections $A^*_\alpha $ form a finite group
$W_{{\frak g}} $ known as the Weyl group. One may introduce in a natural way
an action of the Weyl group on the Cartan-Weyl basis, namely:
\begin{eqnarray*}\label{eq:32.3}
A^*_\alpha (H_\beta ) \equiv A_\alpha
H_\beta A^{-1}_{\alpha } = H_{A^*_\alpha \beta }, \qquad A^*_\alpha
(E_\beta ) \equiv A_\alpha E_\beta A^{-1}_{\alpha } = n_{\alpha ,\beta }
E_{A^*_\alpha \beta }, \quad n_{\alpha ,\beta }=\pm 1.
\end{eqnarray*}
It is also well known that the matrices $A_\alpha  $ are given (up to a
factor from the Cartan subgroup) by
$
A_\alpha ={\rm e}^{E_\alpha } {\rm e}^{-E_{-\alpha }} {\rm e}^{E_\alpha } H_A,
$
where $H_A $ is a conveniently chosen element from the Cartan
subgroup such that $H_A^2=\openone  $. 

\subsection{The Reduction Group}\label{ssec:2.2}

The main idea underlying Mikhailov's reduction group \cite{2} is to impose
algebraic restrictions on the Lax operators $L $ and $M $ which will be
automatically compatible with the corresponding equations of motion.
Due to the purely Lie-algebraic nature of the Lax
representation this is most naturally done by imbedding the
reduction group as a subgroup of $\mbox{Aut}\, {\frak g} $ -- the group of
automorphisms of ${\frak g} $. Obviously, to each reduction imposed on $L $
and $M $ there will correspond a reduction of the space of fundamental
solutions ${\bf S}_\Psi \equiv \{\Psi (x,t,\lambda )\} $ of (\ref{eq:1.5}).

Some of the simplest ${\mathbb Z}_2 $-reductions of Zakharov--Shabat
systems have been known for a long time (see \cite{2}) and are related to
outer automorphisms of ${\frak g} $ and ${\frak G} $, namely:
\begin{eqnarray}\label{eq:C-1}
C_1\left( \Psi (x,t,\lambda ) \right) =  A_1 \Psi^\dag (x,t,\kappa
(\lambda )) A_1^{-1} = \tilde{\Psi}^{-1}(x,t,\lambda ),
\qquad \kappa (\lambda )=\pm \lambda ^*,
\\
C_2\left( \Psi (x,t,\lambda ) \right) =  A_3 \Psi^* (x,t,\kappa
(\lambda )) A_3^{-1} = \tilde{\Psi}(x,t,\lambda ),
\end{eqnarray}
where $A_1$ and $A_3$ are elements of the group of authomorphisms $\mbox{Aut}\,{\frak g}$
of the algebra ${\frak g}$.
Since our aim is to preserve the form of the Lax pair, we limit ourselves
by automorphisms preserving the Cartan subalgebra ${\frak  h} $. 
The reduction group $G_R $ is a finite group which preserves the
Lax representation, i.e. it ensures that the reduction
constraints are automatically compatible with the evolution. $G_R $ must
have two realizations:

i) $G_R \subset {\rm Aut}{\frak g} $

ii) $G_R
\subset {\rm Conf}\, \Bbb C $, i.e. as conformal mappings of the complex
$\lambda $-plane.

To each $g_k\in G_R $ we relate a reduction
condition for the Lax pair as follows \cite{2}:
\begin{equation}\label{eq:2.1}
C_k(U(\Gamma _k(\lambda ))) = \eta_k U(\lambda ),
\end{equation}
where $U(x,\lambda )=q(x)-\lambda J $, $C_k\in \mbox{Aut}\; {\frak g} $ and
$\Gamma _k(\lambda )$ are the images of $g_k $ and $\eta_k =1 $ or $-1 $
depending on the choice of $C_k $.  Since $G_R $ is a finite group then
for each $g_k $ there exist an integer $N_k $ such that $g_k^{N_k}
=\openone $. 

It is well known that $\Aut {\frak g} \equiv V\otimes \Aut_0 {\frak g} $ where
$V $ is the group of outer automorphisms (the symmetry group of the Dynkin
diagram) and $\Aut_0 {\frak g} $ is the group of inner automorphisms. Since
we start with $I,J\in {\frak h} $ it is natural to consider only those inner
automorphisms that preserve the Cartan subalgebra ${\frak h} $. Then $\Aut_0
{\frak g} \simeq \Ad_H \otimes W $ where $\Ad_H $ is the group of similarity
transformations with elements from the Cartan subgroup 
and $W $ is the Weyl group of ${\frak g} $. 

Generically each element $g_k\in G $ maps $\lambda  $ into a
fraction-linear function of $\lambda  $. Such action however is
appropriate for a more general class of Lax operators which are fraction
linear functions of $\lambda  $.

\section{The Caudrey--Beals--Coifman systems}\label{sec:3}

\subsection{Fundamental analytical solutions and scattering data for real $J
$.}
\label{3.1}

The direct scattering problem for the Lax operator (\ref{eq:1.5}) is based
on the Jost solutions:
\begin{eqnarray}\label{eq:3.1.1}
\lim_{x \to \infty }\psi (x,\lambda ){\rm e}^{{\rm i}\lambda Jx}=\openone , \qquad
\lim_{x \to -\infty }\phi (x,\lambda ){\rm e}^{{\rm i}\lambda Jx}=\openone ,
\end{eqnarray}
and the scattering matrix
\begin{eqnarray}\label{eq:3.1.2}
T(\lambda )=(\psi (x,\lambda ))^{-1}\phi (x,\lambda ).
\end{eqnarray}
The fundamental analytic solutions (FAS) $\chi^{\pm} (x,\lambda ) $ of
$L(\lambda ) $ are analytic functions of $\lambda  $ for
$\mbox{Im}\,\lambda \gtrless 0$ and are related to the Jost solutions by
\cite{G*86}
\begin{eqnarray}\label{eq:3.1.3}
\chi ^{\pm}(x,\lambda )=\phi (x,\lambda )S^{\pm}(\lambda )=
\psi ^{\pm}(x,\lambda )T^{\mp}(\lambda )D^{\pm}(\lambda ),
\end{eqnarray}
where $T^{\pm}(\lambda ) $, $S^{\pm}(\lambda ) $ and $D^{\pm}(\lambda ) $
are the factors of the Gauss decomposition of the scattering matrix:
\begin{eqnarray}\label{eq:3.1.4}
&&T(\lambda )=T^-(\lambda )D^+(\lambda )\hat{S}^{+}(\lambda )=
T^+(\lambda )D^-(\lambda )\hat{S}^-(\lambda ) \\
&&T^{\pm}(\lambda )=\exp \left(\sum_{\alpha >0}t^{\pm}_{\pm \alpha
}(\lambda )E_{\alpha } \right), \qquad
S^{\pm}(\lambda )=\exp \left(\sum_{\alpha >0}s^{\pm}_{\pm \alpha
}(\lambda )E_{\alpha } \right),  \nonumber\\
&&D^{+}(\lambda )=I\exp \left(\sum_{j=1}^{r}{2d^+(\lambda )  \over (\alpha
_j,\alpha _j)}H_j \right), \qquad
D^{-}(\lambda )=I\exp \left(\sum_{j=1}^{r}{2d^-(\lambda )  \over (\alpha
_j,\alpha _j)}H_j^- \right). \nonumber
\end{eqnarray}
Here $H_j=H_{\alpha _j} $, $H_j^-=w_0(H_j) $, $\hat{S}\equiv S^{-1} $, I
is an element from the universal center of the corresponding Lie group
${\frak  G} $ and the superscript $+ $ (or $- $) in the Gauss factors
means upper- (or lower-) triangularity for $T^{\pm} (\lambda )  $, $S^{\pm}
(\lambda )  $  and shows that $D^{+} (\lambda )  $ (or $D^{-} (\lambda )
$) are analytic functions with respect to $\lambda  $ for $\mbox{Im}\,
\lambda  >0 $ (or $\mbox{Im}\, \lambda <0 $ respectively).

On the real axis $\chi ^+(x,\lambda )$ and $\chi ^-(x,\lambda ) $
are linearly related by:
\begin{eqnarray}\label{eq:3.1.6}
\chi ^+(x,\lambda )=\chi ^-(x,\lambda )G_0(\lambda ), \qquad
G_0(\lambda )=S^+(\lambda )\hat{S}^-(\lambda ),
\end{eqnarray}
and the sewing function $G_0(\lambda ) $ may be considered as a minimal
system of scattering data provided the Lax operator (\ref{eq:1.5}) has no
discrete eigenvalues \cite{G*86}.

\subsection{The CBC Construction for Semisimple Lie Algebras}
\label{3.2}

Here we will sketch the construction of the FAS for the case of
complex-valued regular Cartan element $J$: $\alpha (\psi )\neq 0$,
following the general ideas of Beals and Coifmal \cite{BC} for
the $sl(n) $ algebras and \cite{VYa} for the orthogonal and
symplectic algebras. These ideas consist of the following:
\begin{enumerate}

\item For potentials $q(x) $ with small norm $||q(x)||_{L^1} <1$ one can
divide the complex $\lambda  $--plane into sectors and then construct an
unique FAS $m_{\nu }(x,\lambda ) $ which is analytic in each of these
sectors $\Omega _{\nu } $;

\item For these FAS in each sector there is a certain Gauss decomposition
problem for the scattering matrix $T(\lambda ) $ which has an unique
solution in the case of absence of discrete eigenvalues.

\end{enumerate}
The main difference between the cases of real-valued and complex-valued $
J $ lies in the fact that for complex $J $ the Jost solutions and the
scattering data exist only for the potentials on compact support.

We define the regions (sectors) $\Omega _{\nu } $ as consisting of those $
\lambda  $'s for which $\mbox{Im}\, (\lambda \alpha (J))\neq 0 $ for any $
\alpha \in \Delta  $. Thus the boundaries of the $\Omega _{\nu } $'s
consist of the set of straight lines:
\begin{eqnarray}\label{eq:3.2.1}
l_{\alpha }\equiv \{ \lambda : \mbox{Im}\, \lambda \alpha (J)=0, \qquad
\alpha \in \Delta \},
\end{eqnarray}
and to each root $\alpha  $ we can associate a certain line
$l_{\alpha } $; different roots may define coinciding lines.

Note that with the change from $\lambda  $ to $\lambda {\rm e}^{{\rm i}\eta } $
and $ J $ to $J{\rm e}^{-{\rm i}\eta } $ (this leads the product $\lambda \alpha
(J) $ invariant) we can always choose $l_1 $ to be along the positive real
$\lambda  $ axis.

To introduce an ordering in each sector $\Omega _{\nu } $ we choose the
vector $\vec{a}_{\nu }(\lambda )\in {\mathbb E}^r $ to be dual to the
element
$\mbox{Im}\, \lambda J \in {\frak  h} $. Then in each sector we split
$\Delta  $ into
\begin{eqnarray}\label{eq:3.2.2}
\Delta =\Delta _{\nu }^+ \cup \Delta _{\nu }^-, \qquad
\Delta _{\nu }^{\pm}=\{\alpha \in \Delta : \mbox{Im}\, \lambda \alpha
(J)\gtrless 0,\, \lambda \in \Omega _{\nu }\}.
\end{eqnarray}
If $\lambda \in \Omega _{\nu } $ then $-\lambda \in \Omega _{M+\nu } $ (if
the lines $l_{\alpha } $ split the complex $\lambda  $-plane into $2M $
sectors). We need also the subset of roots:
\begin{eqnarray}\label{eq:3.2.2b}
\delta _{\nu }=\{\alpha \in \Delta \, : \, \mbox{Im}\, \lambda \alpha
(J)=0, \, \lambda \in l_{\nu }\}
\end{eqnarray}
which will be a root system of some subalgebra ${\frak  g}_{\nu }\subset
{\frak  g} $. Then we can write that
\begin{eqnarray}\label{eq:3.2.3}
{\frak  g}= \mathop{\oplus}\limits_{\nu =1}^{M} {\frak  g}_{\nu }  \qquad
\Delta = \mathop{\cup}\limits_{\nu =1}^{M}\delta _\nu  \qquad
\delta _{\nu }=\delta _{\nu }^+ \cup \delta _{\nu }^-, \qquad
\delta _{\nu }^{\pm}=\delta _{\nu }\cap \Delta _{\nu }^\pm \nonumber .
\end{eqnarray}
Thus we can describe in more details the sets $\Delta _{\nu }^{\pm} $:
\begin{eqnarray}\label{eq:3.2.4}
\Delta _k^+=\delta _1^+ \cup \delta _2^+ \cup \dots \cup \delta _k^+
\cup \delta _{k+1}^-\cup \dots \cup \delta _M^-, \quad \Delta _{k+M}^+=\Delta _k^-, \quad
k=1,\dots , M.
\end{eqnarray}
Note that each ordering in $\Delta  $ can be obtained from the
"canonical" one by an action of a properly chosen element of the
weyl group ${\frak W}({\frak  g}) $.

Now in each sector $\Omega _{\nu } $ we introduce the FAS $\chi _{\nu
}(x,\lambda ) $ and $m_{\nu }(x,\lambda )=\chi _{\nu }(x,\lambda
){\rm e}^{{\rm i}\lambda Jx} $ satisfying the equivalent equation:
\begin{eqnarray}\label{eq:3.2.5}
{\rm i}{{\rm d} m_{\nu }  \over {\rm d} x} + q(x)m_{\nu }(x,\lambda )-\lambda
[J,m_{\nu }(x,\lambda )]=0, \qquad  \lambda \in \Omega _{\nu }.
\end{eqnarray}
If $q(x) $ is a potential on compact support then the FAS $m_{\nu
}(x,\lambda ) $ are related to the Jost solutions by
\begin{eqnarray}\label{eq:3.2.6}
&&m_{\nu }(x,\lambda )=\phi (x,\lambda )S_{\nu }^+(\lambda ){\rm e}^{{\rm i}
\lambda Jx} =\psi (x,\lambda )T_{\nu }^-(x,\lambda )D_{\nu }^+(\lambda
){\rm e}^{{\rm i}\lambda Jx}, \\
&&m_{\nu -1}(x,\lambda )=\phi (x,\lambda )S_{\nu
}^-(\lambda ){\rm e}^{{\rm i}\lambda Jx} =\psi (x,\lambda )T_{\nu }^+(x,\lambda
)D_{\nu }^-(\lambda ){\rm e}^{{\rm i}\lambda Jx}, \qquad \lambda \in l_{\nu }.
\nonumber
\end{eqnarray}
From the definitions of $m_{\nu }(x,\lambda ) $
and the scattering matrix $T(\lambda ) $ we have
\begin{eqnarray}\label{eq:3.2.7}
T(\lambda )=T_{\nu }^-(\lambda )D_{\nu }^+(\lambda )\hat{S}_{\nu }^+
(\lambda )= T_{\nu }^+(\lambda )D_{\nu }^-(\lambda )\hat{S}_{\nu }^-
(\lambda ) , \quad \lambda \in l_{\nu }
\end{eqnarray}
where in the first equality we take $\lambda =\mu {\rm e}^{{\rm i}0} $ and for the
second-- $\lambda =\mu {\rm e}^{-{\rm i}0} $ with $\mu \in l_{\nu } $. The
corresponding expressions for the Gauss factors have the form:
\begin{eqnarray}\label{eq:3.2.8}
&&S_{\nu }^+(\lambda )=\exp \left(\sum_{\alpha \in \Delta _{\nu
}^+}s_{\nu ,\alpha }^+(\lambda )E_{\alpha } \right), \qquad
S_{\nu }^-(\lambda )=\exp \left(\sum_{\alpha \in \Delta _{\nu
-1}^+}s_{\nu ,\alpha }^-(\lambda )E_{-\alpha } \right),  \nonumber\\
&&T_{\nu }^+(\lambda )=\exp \left(\sum_{\alpha \in \Delta _{\nu
-1}^+}t_{\nu ,\alpha }^+(\lambda )E_{\alpha } \right), \qquad
T_{\nu }^-(\lambda )=\exp \left(\sum_{\alpha \in \Delta _{\nu
}^+}t_{\nu ,\alpha }^-(\lambda )E_{-\alpha } \right),  \nonumber\\
&&D_{\nu }^+(\lambda )=\exp ({\bf d}_{\nu }^+(\lambda )\cdot {\bf
H}_{\nu }), \qquad
D_{\nu }^-(\lambda )=\exp ({\bf d}_{\nu }^-(\lambda )\cdot {\bf
H}_{\nu -1}).
\end{eqnarray}
Here ${\bf d}_{\nu }^{\pm}(\lambda )=(d_{\nu ,1}^{\pm},\dots ,d_{\nu
,r}^{\pm}) $ is a vector in the root space and
\begin{eqnarray}\label{eq:3.2.9}
{\bf H}_{\eta}= \left( {2H_{\eta ,1}  \over (\alpha _{\eta,1},
\alpha _{\eta,1}) }, \dots , {2H_{\eta ,r}  \over (\alpha _{\eta,r},
\alpha_{\eta,r}) } \right), \qquad
({\bf d}_{\nu }^{\pm}(\lambda ), {\bf H}_{\eta})=
\sum_{k=1}^{r}{2d_{\nu ,k}^{\pm}(\lambda )H_{\eta ,k}  \over (\alpha
_{\eta ,k}, \alpha _{\eta ,k}) },
\end{eqnarray}
where $\alpha _{\eta ,k} $ is the $k$-th simple root of ${\frak  g} $ with
respect to the ordering $\Delta _{\eta}^+$  and $H_{\eta ,k} $ are their
dual elements in the Cartan subalgebra ${\frak  h} $.

\section{The Gauge Group Action}\label{sec:4}

\subsection{The class of the gauge equivalent NLEE's}\label{ssec:4.1}

The notion of gauge equivalence allows one to associate to any Lax pair 
of the type (\ref{eq:1.5}), (\ref{eq:1.7})
an
equivalent one \cite{VYa}, solvable by the inverse scattering
method for the gauge equivalent linear problem:
\begin{eqnarray}\label{eq:2.3}
\tilde{L}\tilde{\psi }(x,t,\lambda )\equiv \left({\rm i} {{\rm d} \over {\rm d}
x}-\lambda S \right) \tilde{\psi }(x,t,\lambda )=0, \nonumber\\
\tilde{M}\tilde{\psi }(x,t,\lambda )\equiv \left({\rm i} {{\rm d}  \over
{\rm d} t}-\lambda f(S) \right) \tilde{\psi }(x,t,\lambda)=0,
\end{eqnarray}
where $\tilde{\psi }(x,t,\lambda ) = g^{-1}(x,t)\psi (x,t,\lambda )$,
\begin{eqnarray}\label{eq:2.4}
S = \mbox{Ad}_{g}\cdot J \equiv g^{-1}(x,t)Jg(x,t),
\end{eqnarray}
and $g(x,t)=m_\nu (x,t,0) $ is FAS at $\lambda =0 $. 
The functions $m_\nu (x,t,\lambda)$ are analytic with respect to $\lambda$ 
in each sector $\Omega_\nu$ and do not loose their analyticity for $\lambda=0$
(in the case of potential on compact support). From the integral representation 
for the FAS $m_\nu (x,t,\lambda)$ at $\lambda=0$ it follows that
\[
m_1(x,t,0)= \cdots = m_\nu(x,t,0)=\cdots = m_{2M}(x,t,0).
\]
Therefore the gauge group action is well defined.
The
zero-curvature condition $[\tilde{L},\tilde{M}]=0 $ gives:
\begin{eqnarray}\label{eq:2.5}
S_t- {{\rm d}  \over {\rm d} x }f(S)=0,
\end{eqnarray}
where $f(S)= \sum_{p=0}^{r-1}\alpha _p S^{2p+1} $ is an
odd polynomial of $S $. 
Both Lax operators $L(\lambda)$ and $\tilde{L}(\lambda)$ have equivalent spectral properties 
and spectral data and therefore the classes of NLEE's related to them are equivalent.
It is natural that $f({\cal
S})=g^{-1}(x,t) I g(x,t) $, i.e., it is uniquely determined by $I $. Both
$J $ and $I $ belong to the Cartan subalgebra $\fr{h} $ so they have
common set of eigenspaces.

1) $\fr{g}\simeq {\bf A}_r = sl(n) $ with $n=r+1 $. We have
\[
J= \mbox{diag}\, (J_1,\dots, J_n) , \qquad I= \mbox{diag}\,  (I_1,\dots, I_n),
\]
and the only constraint on the eigenvalues $J_k $ and $I_k $ is $\tr J =
\tr I =0 $. The projectors on the common eigensubspaces of $J $ and $I $
are given by:
\begin{equation}\label{eq:pi_k}
\pi_k (J) = \prod_{s\neq k} {J-J_s  \over J_k-J_s } =
\mbox{diag}\,  (0,\dots, 0,\mathop{1}\limits_k , 0,\dots, 0).
\end{equation}
Next we note that
$I= \sum_{k=1}^{n} I_k \pi_k(J).$ In order to derive $f({\cal  S}) $ for $\fr{g}\simeq sl(n) $ we need to
apply the gauge transformation to (\ref{eq:2.5}) with the result:
\begin{equation}\label{eq:f-sln}
f({\cal  S}) = \sum_{k=1}^{n} I_k \pi_k({\cal  S}),
\end{equation}
i.e., $f({\cal  S})  $ is a polynomial of order $n-1 $. Obviously ${\cal
S} $ is restricted by:
\begin{equation}\label{eq:S-cha}
\prod_{k=1}^{n} ({\cal  S} - J_k) =0, \qquad \tr {\cal  S}^k = \tr J^k,
\end{equation}
for $k=2,\dots,n $.

2) ${\frak g} \simeq {\bf B}_r, {\bf D}_r$ In order to express $f(S) $ through
their eigenvalues $J_k $ and $I_k $ we introduce the
diagonal matrix-valued functions:
\begin{equation}\label{eq:f_k}
f_k(J) = {J  \over J_k } \prod_{s\neq k}^{} {J^2 -J_s^2 \over J_k^2 -
J_s^2 } = H_{e_k} \in \fr{h},
\end{equation}
where by $H_{e_k} $ we denote the element in $\fr{h} $ dual to the basis
vector $e_k $ in the root space of $\fr{g} $. Using (\ref{eq:f_k})
and applying $\mbox{Ad}_g $ we get:
\begin{eqnarray}\label{eq:I}
I= \sum_{k=1}^{r} I_k f_k(J),\qquad
f(S)\equiv  g^{-1}(x,t) I g(x,t) = \sum_{k=1}^{r} I_k
f_k(S).
\end{eqnarray}
In addition $S(x,t) $ satisfies the characteristic equations:
\begin{eqnarray}\label{eq:2.6}
S^{\kappa _0}\prod_{k=1}^{r}(S^2-J_k^2)=0,
\end{eqnarray}
where $\kappa _0= 0$ if $ {\frak  g} \simeq C_r$ or $D_r$ and $\kappa _0=1
$, if ${\frak  g} \simeq B_r$.

Then the equation gauge equivalent to (\ref{eq:1.5}) becomes:
\begin{eqnarray}\label{eq:2.8}
S_t-\alpha _0S_x-\sum_{p=1}^{r-1}\alpha _p( S^{2p+1})_x=0.
\end{eqnarray}
The function $S(x,t) \in \fr{g} $ is also subject to constraints;  one of
them is provided by (\ref{eq:2.6}). To construct the others we assume that
$\fr{g} \simeq {\bf B}_r $ or ${\bf D}_r $ and use the typical representation of
$\fr{g} $. It this settings we easily see that all odd powers of $H_{e_k}
$ also belong to the Cartan subalgebra  $\fr{h} $. Thus we conclude that
all odd powers of $S $ also belong to $\fr{g} $. The invariance properties
of the trace lead to:
\begin{equation}\label{eq:tr_Jk}
\tr (J^{2k}) \equiv 2 \sum_{k=1}^{r} J_k^{2k} = \tr (S)^{2k},
\end{equation}
for $k=1,\dots,r $. The conditions (\ref{eq:tr_Jk}) are precisely
$r $ independent algebraic constraints on $S $. Solving
for them we conclude that the number of independent coefficients
in $S $ is equal to the number of roots $|\Delta | $ of
$\fr{g} $.

\subsection{The Minimal Set of Scattering Data for $L(\lambda)$ and $\tilde{L}(\lambda)$}\label{ssec:4.2}

We skip the details about CBC construction which can be found in
\cite{VYa} and go to the minimal set of scattering data for the
case of complex $J $ which are defined by the sets ${\cal  F}_1 $
and ${\cal  F}_2 $ as follows:
\begin{eqnarray}\label{eq:3.2.10}
{\cal  F}_1= \mathop{\cup}\limits_{\nu =1}^{2M}{\cal  F}_{1,\nu }, \qquad
{\cal  F}_2= \mathop{\cup}\limits_{\nu =1}^{2M}{\cal  F}_{2,\nu },
\nonumber\\
{\cal  F}_{1,\nu }= \{\rho _{B,\nu ,\alpha }^{\pm }(\lambda ), \, \alpha
\in \delta _{\nu }^+, \, \lambda \in l_{\nu }\} \qquad
{\cal  F}_{2,\nu }= \{\tau _{B,\nu ,\alpha }^{\pm }(\lambda ), \, \alpha
\in \delta _{\nu }^+, \, \lambda \in l_{\nu }\},
\end{eqnarray}
where
\begin{eqnarray}\label{eq:3.2.12}
\rho _{B,\nu ,\alpha }^{\pm }(\lambda )=\langle S_{\nu }^{\pm}(\lambda
)B\hat{S}_{\nu }^{\pm}(\lambda ), E_{\mp \alpha }\rangle , \qquad
\tau _{B,\nu ,\alpha }^{\pm }(\lambda )=\langle T_{\nu }^{\pm}(\lambda
)B\hat{T}_{\nu }^{\pm}(\lambda ), E_{\mp \alpha }\rangle ,
\end{eqnarray}
with $\alpha \in \delta _{\nu }^+ $, $\lambda \in l_{\nu } $ and
$B $ is a properly chosen regular element of the Cartan
subalgebra ${\frak  h} $. Without loose of generality we can take
in (\ref{eq:3.2.12}) $B=H_{\alpha } $.
Note that the functions $\rho _{B,\nu ,\alpha }^{\pm}(\lambda ) $ and
$\tau _{B,\nu ,\alpha }^{\pm}(\lambda ) $ are continuous functions of
$\lambda  $ for $\lambda \in l_{\nu } $.

If we choose $J $ in such way that $2M=|\Delta | $-- the number of the
roots of ${\frak  g} $. then to each pair of roots $\{\alpha ,-\alpha \} $
one can relate a separate pair of rays $\{l_{\alpha }, l_{\alpha +M}\} $,
and $l_{\alpha }\neq l_{\beta } $ if $\alpha \neq \pm \beta  $. In this
case each of the subalgebras ${\frak  g}_{\alpha } $ will be isomorphic to
$sl(2) $.

In order to determine the scattering data for the gauge equivalent
equations we need to start with the FAS for these systems:
\begin{eqnarray}\label{eq:3.1}
\tilde{m }_\nu^{\pm}(x,\lambda )=g^{-1}(x,t)m_\nu^{\pm}(x,\lambda )g_-,
\end{eqnarray}
where $g_-= \lim_{x \to -\infty }g (x,t) $ and due to (\ref{eq:1.7}) and
$g_-=\hat{T}(0)$.  In order to ensure that the functions $\tilde{\xi
}^{\pm}(x,\lambda ) $ are analytic with respect to $\lambda $ the
scattering matrix $T(0) $ at $ \lambda =0 $ must belong to the
corresponding Cartan subgroup ${\frak  H} $. Then Equation (\ref{eq:3.1})
provide the fundamental analytic solutions of $\tilde{L} $.  We can
calculate their asymptotics for $x\to\pm\infty  $ and thus establish the
relations between the scattering matrices of the two systems:
\begin{eqnarray}\label{eq:3.2}
\lim_{x \to -\infty }\tilde{\xi }^+(x,\lambda )=
e^{-{\rm i} \lambda Jx}T(0)S^+(\lambda )\hat{T}(0) \qquad
\lim_{x \to \infty }\tilde{\xi }^+(x,\lambda )=
e^{-{\rm i} \lambda Jx}T^-(\lambda )D^+(\lambda )\hat{T}(0)
\end{eqnarray}
with the result:
$
\tilde{T}(\lambda )= T(\lambda )\hat{T}(0)$.
Obviously  $\tilde{T}(0)=\openone  $.
The factors in the corresponding Gauss decompositions are related by:
\begin{eqnarray}\label{eq:3.4}
\tilde{S}^{\pm}(\lambda )= T(0)S^{\pm}(\lambda )\hat{T}(0), \qquad
\tilde{T}^{\pm}(\lambda )=T^{\pm}(\lambda ) \nonumber\qquad
\tilde{D}^{\pm}(\lambda )=D^{\pm}(\lambda )\hat{T}(0).
\end{eqnarray}
On the real axis again the FAS $\tilde{\xi }^+(x,\lambda ) $ and
$\tilde{\xi }^-(x,\lambda ) $ are related by $
\tilde{\xi }^+(x,\lambda )=\tilde{\xi }^-(x,\lambda )\tilde{G}_0(\lambda )
$
with the normalization condition $\tilde{\xi }(x,\lambda =0)=\openone $
and $\tilde{G}_0(\lambda )=\tilde{S}^+(\lambda )\hat{\tilde{S}}^-(\lambda
) $ again  can be considered as a minimal set of scattering data.

The minimal set of scattering data for the gauge-equivalent CBC systems  are defined by the sets $\tilde{\cal  F}_1 $
and $\tilde{\cal  F}_2 $ as follows:
\begin{eqnarray}\label{eq:3.2.10a}
\tilde{\cal  F}_1= \mathop{\cup}\limits_{\nu =1}^{2M}\tilde{\cal  F}_{1,\nu }, \qquad
\tilde{\cal  F}_2= \mathop{\cup}\limits_{\nu =1}^{2M}\tilde{\cal  F}_{2,\nu },
\nonumber\\
\tilde{\cal  F}_{1,\nu }= \{\tilde{\rho }_{B,\nu ,\alpha }^{\pm }(\lambda ), \, \alpha
\in \delta _{\nu }^+, \, \lambda \in l_{\nu }\} \qquad
\tilde{\cal  F}_{2,\nu }= \{\tilde{\tau }_{B,\nu ,\alpha }^{\pm }(\lambda ), \, \alpha
\in \delta _{\nu }^+, \, \lambda \in l_{\nu }\},
\end{eqnarray}
where
\begin{eqnarray}\label{eq:3.2.12a}
\tilde{\rho }_{B,\nu ,\alpha }^{\pm }(\lambda )=\langle T(0)S_{\nu }^{\pm}(\lambda
)B\hat{S}_{\nu }^{\pm}(\lambda )\hat{T}(0), E_{\mp \alpha }\rangle , \quad
\tilde{\tau }_{B,\nu ,\alpha }^{\pm }(\lambda )=\langle T_{\nu }^{\pm}(\lambda
)B\hat{T}_{\nu }^{\pm}(\lambda ), E_{\mp \alpha }\rangle ,
\end{eqnarray}
with $\alpha \in \delta _{\nu }^+ $, $\lambda \in l_{\nu } $ and
$B $ is again a properly chosen regular element of the Cartan
subalgebra ${\frak  h} $. Without loose of generality we can take
in (\ref{eq:3.2.12a}) $B=H_{\alpha } $ (as in (\ref{eq:3.2.12})).
That the functions $\tilde{\rho }_{B,\nu ,\alpha }^{\pm}(\lambda ) $ and
$\tilde{\tau }_{B,\nu ,\alpha }^{\pm}(\lambda ) $ are continuous functions of
$\lambda  $ for $\lambda \in l_{\nu } $ and have the same analyticity properties as the 
functions  $\rho _{B,\nu ,\alpha }^{\pm}(\lambda ) $ and
$\tau _{B,\nu ,\alpha }^{\pm}(\lambda ) $.

\subsection{Integrals of Motion and Hierarchies of Hamiltonian Structures}\label{ssec:4.3}

Both classes of NLEE's are infinite dimensional completely integrable Hamiltonian
systems and possess hierarchies of Hamiltonian structures. 

The phase space ${\cal  M}_{\rm CBC} $ is the linear space of
all off-diagonal matrices $q(x,t) $ tending fast enough to zero for
$x\to\pm\infty $. The hierarchy of pair-wise compatible symplectic
structures on ${\cal  M}_{\rm CBC} $ is provided by the $2 $-forms:
\begin{equation}\label{eq:ome-nls}
\Omega _{\rm CBC}^{(k)} = {\rm i } \int_{-\infty }^{\infty } dx \tr
\left( \delta q(x,t) \wedge \Lambda ^k [J, \delta q(x,t) ]
\right),
\end{equation}
where $\Lambda= (\Lambda_+ + \Lambda_-)/2
$ is the generating (recursion) operator for (\ref{eq:1.5}) defined as follows:
\[
\Lambda _\pm Z(x) =\mbox{ad}_J^{-1}(1 - \pi_0)\left( i {dZ \over dx}
+ [q(x), Z(x)] + i \left[ q(x),  \pi_0 \int_{\pm\infty }^{x} dy\,
[q(y),Z(y)] \right] \right),
\]
where $\pi_0 (X)=\mbox{ad}_J^{-1}\circ \mbox{ad}_J (X)$.
The symplectic forms $\Omega _{\rm CBC}^{(k)}$ can be expressed in terms of 
the scattering data for $L(\lambda)$:
\begin{eqnarray}\label{eq:3.2.a}
&& \Omega _{\rm CBC}^{(k)} = {c_k \over 2\pi }\sum_{\nu =1}^{M} \int_{\lambda \in
l_{\nu }\cup l_{M+\nu } } d\lambda \lambda ^k \left( \Omega
_{0,\nu }^+(\lambda ) - \Omega _{0,\nu }^-(\lambda )\right), \nonumber\\
&& \Omega _{0,\nu }^\pm(\lambda ) = \left\langle \hat{D}_{\nu }
^\pm(\lambda ) \hat{T}_{\nu }^\mp(\lambda ) \delta T_{\nu }^\mp(\lambda )
D_{\nu }^\pm(\lambda ) \wedge \hat{S}_{\nu }^\pm(\lambda ) \delta
S_{\nu }^\pm(\lambda ) \right\rangle .
\end{eqnarray}
Note that the kernels of $\Omega _{\rm CBC}^{(k)}$ differ only by the factor $\lambda^k$ 
so all of them can be casted into canonical form simultaneously.

The phase space ${\cal  M}_{\rm gauge} $ of the gauge equivalent to the CBC systems 
is the manifold of all
${\cal S}(x,t) $ determined by the second relation in (\ref{eq:2.4}). The
family of compatible $2 $-forms is:
\begin{equation}\label{eq:ome-hf}
\tilde{\Omega} _{\rm gauge}^{(k)} = {i \over 4} \int_{-\infty }^{\infty }
dx \tr \left( \delta S^{(0)} \wedge \tilde{\Lambda} ^k [S^{(0)}, \delta
S^{(0)}(x,t) ] \right).
\end{equation}
Here $\tilde{\Lambda } $ is the recursion
operator for the gauge equivalent to the CBC systems:
\[
\widetilde{\Lambda }_\pm \widetilde{Z} = i\mbox{ad}_{S(x)}^{-1} \left(1-\widetilde{\pi}_0(x) \right) \left\{ {d\widetilde{Z}  \over dx }  +
\sum_{k=1}^{2} [\widetilde{h}_k(x), \mbox{ad}_{\mathcal{S}(x)}^{-1} ]
\int_{\pm\infty }^{x} dy\, \left\langle [\widetilde{h}_k(y),
\mbox{ad}_{\mathcal{S}(y)}^{-1} \mathcal{S}_y], \widetilde{Z}(y) \right\rangle,
\right\}
\]
where $\widetilde{h}_k(x,t)=g^{-1}(x,t)H_kg(x,t)$, and $\langle
H_k,H_j\rangle = \langle \widetilde{h}_k(x,t),\widetilde{h}_j(x,t)\rangle
=\delta_{jk}$. 

The spectral theory of these two operators $\Lambda$ and $\tilde{\Lambda }$ underlie all the fundamental
properties of these two classes of gauge equivalent NLEE, for details see
\cite{VYa}.  Note that the gauge transformation relates nontrivially the
symplectic structures, i.e. $\Omega _{\rm NLSE}^{(k)} \simeq\tilde{\Omega}
_{\rm HFE}^{(k+2)} $ \cite{RST,VYa}. 

\section{Conclusions}
We will finish this article with several concluding remarks.
To CBC systems and their gauge equivalent one can apply the analysis \cite{VYa} and derive the completeness relations for the corresponding system of "squared" solutions. Such analysis will allow one to prove the pair-wise compatibility of the Hamiltonian structures and eventually to derive their action-angle variables, see e.g. \cite{ZM} and \cite{BS} for the ${\bf A}_r$-series.

For the case of singular  $J$ ($\alpha(J)=0$)  the construction of FAS $m_\nu (x,t,\lambda)$   and  $\tilde{m}_\nu (x,t,\lambda)$ requires the use of generalized Gauss decomposition in which the factors $D_\nu^\pm (\lambda)$ are block-diagonal, while $T_\nu^\pm (t,\lambda)$ and  $S_\nu^\pm (t,\lambda)$  are block-triangular. This will be addressed to a subsequent paper.

The approach presented here allows one to consider CBC systems with more general    $\lambda$- dependence, like the Principal Chiral field models and other relativistic invariant fiels theories \cite{Za*Mi}.

If   ${\frak g}\simeq so(5)$ then the corresponding gauge equivalent system describes isoparametric surfaces \cite{Ferap}.

Finally, some open problems are :
1)to study the internal structure of the soliton solutions and soliton interactions (for both types of systems);
2) to study reductions of the gauge equivalent to CBC systems.

\section*{Acknowledgements}

We thank professors E. V. Ferapontov, V. S. Gerdjikov, D. J. Kaup and A. V. Mikhailov for the numerous stimulating discussions. One of us (GGG) thanks the organizing committee of the NEEDS-2007 conference for the scholarship provided and for the
warm hospitality in Ametlla de Mar.  The support by the National Science
Foundation of Bulgaria, contract No. F-1410 and by the Science Foundation of Ireland is acknowledged.

\label{lastpage}


\begin{thebibliography}{99}
\small

\bibitem{AKNS}
   Ablowitz M J, Kaup D J, Newell A C, Segur H., The inverse
scattering transform -- Fourier analysis for nonlinear problems, {\it  Studies
   in Appl.\ Math.}\ {\bf 53} (1974), n 4, 249--315.



\bibitem{BC}
Beals R., Coifman R R., Scattering and inverse scattering for
first order systems.,  {\it Commun.\ Pure and Appl.\ Math.}\ {\bf 37} (1984),
n 1, 39--90.

\bibitem{BC2} Beals  R.,  Coifman R. R., {\it  Commun. Pure \& Appl. Math.}
{\bf 38} (1985), 29--42.

\bibitem{BC3} Beals R., Coifman R. R., Scattering and inverse
scattering for first order systems II, {\it  Inverse Problems} {\bf 3}  (1987),
577--594.\\
Beals R., Coifman R. R., Linear spectral problems, nonlinear
equations and the delta-method, {\it  Inverse Problems} {\bf 5} (1989), 87--130.



\bibitem{BS}
Beals R., Sattinger D.,  On the complete integrability of completely
integrable systems, {\it  Commun. Math. Phys.} {\bf 138} (1991), 409--436.


\bibitem{LA} Bourbaki N.,
{\it Elements de mathematique. Groupes et algebres de Lie. Chapters
I--VIII}, Hermann, Paris, 1960--1975.



\bibitem{Caud*80} Caudrey P. J., The inverse problem for a general
$N\times N $ spectral equation, {\it  Physica D} {\bf D6} (1982), 51-66.


\bibitem{FG} Coxeter H. S. M., Moser W. O. J., {\it Generators and
relations for discrete groups}, Springer Verlag, Berlin Heidelberg New
York, 1972.



\bibitem{FaTa} Faddeev L D., Takhtadjan L A., {\it Hamiltonian approach
in the theory of solitons\/}, Springer Verlag, Berlin, 1987.

\bibitem{Ferap} Ferapontov E. V., Isoparametric hypersurfaces in spheres,
integrable nondiagonalizable systems of hydrodynamic type,
and N-wave systems, {\it Diff. Geom. Appl.} {\bf 5} (1995), 335--369.

\bibitem{ForGib*80b} Fordy A P., Gibbons J., Integrable nonlinear
Klein--Gordon equations and Toda lattices,
{\it  Commun.\ Math.\ Phys. }\ {\bf 77} (1980), 21--30.


\bibitem{ForKu*83}
{} Fordy A P., Kulish P P., Nonlinear Schr\"{o}dinger equations and
simple Lie algebras, {\it  Commun.\ Math.\ Phys.}\ {\bf 89} (1983), 427--443.



\bibitem{VSG*94} Gerdjikov V. S.,    The Generalized Zakharov--Shabat
System and the Soliton Perturbations, {\it  Teor. Mat. Fiz. }{\bf 99} (1994),
292-299.



\bibitem{G*86} Gerdjikov V S., Kulish P P., 
The generating operator for the $n\times n$ linear system,
{\it  Physica D} {\bf 3} (1981), 549--564. \\
Gerdjikov V S., Generalized Fourier transforms  for the soliton
equations. Gauge covariant formulation, {\it  Inverse Problems} {\bf 2}  (1986),
51--74.

\bibitem{vgn}
 Gerdjikov V. S., Grahovski G. G., Kostov N. A.,
Reductions of $N $-wave interactions related to simple Lie algebras I:
${\bf Z}_2$- reductions, {\it  J. Phys. A: Math. and Gen.} {\bf 34}
(2001), 9425--9461.

\bibitem{vgn1}
 Gerdjikov V. S., Grahovski G. G., Kostov N. A.,
On N-wave type systems and their
gauge equivalent, {\it  Eur. Phys. J.} B {\bf 29} (2002), 243–248.

\bibitem{vgrn}
 Gerdjikov V. S., Grahovski G. G., Ivanov R. I., Kostov N. A.,
$N $-wave interactions related to simple Lie algebras.
${\bf Z}_2$- reductions and soliton solutions, {\it  Inverse Problems }{\bf 17}
(2001), 999--1015.

\bibitem{VYa}
Gerdjikov V S., Yanovski A B., 
Completeness of the eigenfunctions for the Caudrey -- Beals -- Coifman
system, {\it  J. Math. Phys.} {\bf 35} (1994), 3687--3725.

\bibitem{g} Grahovski G. G., On The Reductions and Scattering Data for the CBC System., In
"Geometry, Integrability and Quantization III", Eds: I. Mladenov and G. Naber, Coral
Press, Sofia, 2002, pp.262--277.


\bibitem{GoGr} Goto M., Grosshans F., {\it Semisimple Lie algebras},
Lecture Notes in Pure and Applied Mathematics vol. {\bf 38}, M.Dekker
Inc., New York \& Basel, 1978.




\bibitem{Helg} Helgasson S.,
{\it Differential geometry, Lie groups and symmetric spaces},
Academic Press, 1978.

\bibitem{Hu} Humphreys J. E., {\it Reflection Groups and Coxeter Groups},
Cambridge University Press, Cambridge, 1990.

\bibitem{Ivanov}  Ivanov R. I., On the dressing method for the generalised Zakharov-Shabat system, 
{\it Nuclear Physics B} {\bf 694} (2004), 509--524.


\bibitem{2} Mikhailov A V.,   The reduction problem and the inverse
scattering problem, {\it  Physica D} {\bf 3} (1981), 73--117.

\bibitem{MiOlPer*81} Mikhailov A. V. , Olshanetzky M. A., Perelomov A. M.,
Two--dimensional generalized Toda lattice,  {\it  Commun.\ Math.\ Phys.}
{\bf 79} (1981), 473--490.




\bibitem{RST}Reyman A. G., An unified Hamiltonian system on
polynomial bundles and the structure of the stationary problems, {\it  Zap.
Nauch. Semin. LOMI} {\bf 131} (1983), 118--127;  (In Russian).




\bibitem{ZM}
Zakharov V E., Manakov S V.,  {\it Exact theory of resonant interaction of
wave packets in nonlinear media}, INF preprint 74-41, Novosibirsk (1975)
(In Russian).


\bibitem{1} Zakharov V E., Manakov S V., Novikov S P.,
Pitaevskii L I.,
{\it Theory of solitons. The inverse scattering method},
Plenum, N.Y., 1984.


\bibitem{Za*Mi} Zakharov  V. E., Mikhailov A. V.,  On the
integrability of classical spinor models in two--dimensional space--time,
{\it  Commun. Math. Phys.} {\bf 74} (1980), 21--40.



\bibitem{Za*Sh}  Zakharov V. E. and Shabat A. B.,
Exact theory of two-dimensional selffocusing and one-dimensional
automodulation in nonlinear media, {\it  Zh. Exp. Teor. Fiz.} {\bf 61} (1971), 118--134.

\bibitem{Zhou} Zhou X., Direct and inverse scattering transform
with arbitrary spectral singularities, {\it  Commun.  Pure \& Appl.  Math.}
{\bf 42} (1989), 895--938.

\bibitem{ZaTa}  Zakharov V. E.,  Takhtajan L. A.
The equivalence between the nonlinear Schr\"odinger equation and the
Heisenberg ferromagnet equation,
{\it   Teor. Mat. Fiz.} {\bf 38} (1979), 26--35.

\end{thebibliography}
\end{document}